\documentclass{amsart}
\usepackage{graphicx}
\usepackage{url}
\usepackage{color}
\usepackage{subfig}
\usepackage[pdftex]{hyperref}
\usepackage{multirow}
\usepackage{footmisc}
\usepackage{ws-rotating}
\usepackage[utf8]{inputenc}

\def\pprw{8.5in}
\def\pprh{11in}

\def\etal{{\it et al.}}

\definecolor{linkColor}{RGB}{6,125,233}

\begin{document}

\title{Crowdsourcing Health Labels:\\ Inferring Body Weight from Profile Pictures*}\thanks{*This is a preprint of an article appearing at ACM DigitalHealth 2016.}

\author{
Ingmar Weber \and
Yelena Mejova \\
Qatar Computing Research Institute, HBKU\\
Doha, Qatar\\
       \{iweber,ymejova\}@qf.org.qa
}

\maketitle

\begin{abstract}
To use social media for health-related analysis, one key step is the detection of health-related labels for users. But unlike transient conditions like flu, social media users are less vocal about chronic conditions such as obesity, as users might not tweet ``I'm still overweight''. As, however, obesity-related conditions such as diabetes, heart disease, osteoarthritis, and even cancer are on the rise, this obese-or-not label could be one of the most useful for studies in public health.

In this paper we investigate the feasibility of using \emph{profile pictures} to infer if a user is overweight or not. We show that this is indeed possible and further show that the fraction of labeled-as-overweight users is higher in U.S.\ counties with higher obesity rates. Going from public to individual health analysis, we then find differences both in behavior and social networks, for example finding users labeled as overweight to have fewer followers. 
\end{abstract}



\section{Introduction}\label{sec:introduction}

In many developed countries, obesity\footnote{For medical and statistical purposes, ``obese'' is defined as having a Body-Mass Index (BMI) of $\ge 30$, with ``overweight'' referring to BMI $\ge 25$.\label{ftnote:obesity}} levels have reached epidemic proportions. In the U.S.\ alone, medical costs linked to obesity were estimated to be \$147 billion in 2008\footnote{\url{http://www.cdc.gov/chronicdisease/overview/}}, 
such that lower obesity rates could produce productivity gains of \$254 billion.\footnote{\url{http://www.milkeninstitute.org/publications/view/321}} 
In this paper we look at whether public social media data, Twitter in this case, could be used to identify individuals likely to be overweight. Traditional methods utilizing textual self-disclosure suffer from low recall and introduce bias in the character of the detected users. However, profile pictures may convey similar health information in a kind of non-verbal self-disclosure, with a potentially vastly greater coverage.

%
To assess the feasibility of using profile images to infer a user's weight, we ask the following research questions:

\noindent Q1: Is it possible to infer whether an individual is overweight from the profile picture on social media?

\noindent Q2: Is the distribution of users identified to be overweight aligned with county-level offline statistics? 

\noindent Q3: Correcting for the region they are in, are there differences in how users identified to be overweight and not-overweight use social media?

Though the answers to Q1 and Q2 might appear obvious, they are not. Previous work \cite{abbaretal15chi} mentioned a negative result concerning Q1. Instead, we apply a two-stage crowdsourced process to the task. Furthermore, as high obesity areas are typically rural with different social media usage patterns, the answer to Q2 is also not obvious, especially as heavily overweight users might choose not to share any profile image. 
We find a positive answer to all three questions, paving the way for future studies looking at individual-level effects and behavior related to obesity.

\section{Related Work}\label{sec:related}

Famously, Google Flu Trends \cite{ginsberg2009detecting} has inspired researchers to look at Twitter data to monitor flu epidemics \cite{aramaki2011twitter,lampos2010tracking}. Similarly, obesity has been tracked in populations at as small as county-level scales \cite{abbaretal15chi,culotta14chi}. However, it is impossible to validate their predictions at the individual level. Our research, thus, allows us to juxtapose overweight users in high obesity areas with non-overweight users in the same area (see User Analysis Section).

A scalable alternative is computer vision research. Recently, researchers have looked at inferring a person's BMI from images \cite{wenguo13ivc} and at the effect that changes in body weight have on face recognition \cite{wenetal14ijcb}. 
Though the noise in social media image data limits the applicability of these automated techniques.

Though obesity is the most common health problem related to food intake, other researchers have looked at \emph{anorexia} and how it is discussed on social media \cite{syed2013misleading,yom2012pro}. Conceptually, our profile photo based approach might also be able to detect users who are underweight, though we did not experiment with this labeling task. Furthermore, \cite{puhl2001bias} describe the stigmatization and discrimination by body-weight in terms of employment, education, and health care. The individual-level detection of body weight on social networks would help in the monitoring of weight-based bias in interactions captured by the social media.

\section{Data}\label{sec:data}

For our study we opted to focus on counties with extreme levels of obesity, both high and low. Using County Health Rankings\footnote{\url{http://www.countyhealthrankings.org/}}, we selected the 73 most obese counties ($\ge 38.5\%$ obese) and the 43 least obese counties ($< 19.5\%$) for a total of 116 counties (see Table~\ref{tab:county_obesity}).


\begin{table}[ht]
\centering
\caption{The top five most and least obese U.S.\ counties. 
\label{tab:county_obesity}}{
\vspace{-0.4cm}
\begin{tabular}{cccc}
\multicolumn{2}{c}{High Obesity} & \multicolumn{2}{c}{Low Obesity}  \\ \hline
Greene, AL & 48 &        Routt, CO & 14 \\
Jefferson, MS & 45 &     Teton, WY & 14 \\
Lowndes, AL & 45 &       Eagle, CO & 14 \\
Ziebach, SD & 44 &       Santa Fe, NM & 14 \\
Coahoma, MS & 44 &       Pitkin, CO & 14 \\
\end{tabular}}
\end{table}

We then used a high-precision (low-recall) approach to create a sample of users from these counties. 
For each of these 116 counties we then used Followerwonk\footnote{\url{https://followerwonk.com/bio/}} to search for Twitter users whose location strings contained our target locations such as ``Greene, AL''. This approach to ``find a Twitter user from location X'' proved to have high precision upon manual inspection. Still, we removed five counties such as ``Park, CO'' as ``Park'' not only matched Park County but also other cities containing park in their name. The other cases where ``Grand, CO'', ``Lake, CO'', ``Mississippi, AR'' and ``New York, NY''. Note that a low recall was not much of an issue for as we were not interested in finding \emph{all} Twitter users from a given county, but merely a subset.


In addition to the string matching on the location, we required a minimum of 10 tweets, followers and friends, as well as a maximum of 1,000 followers and friends, and a maximum of 5,000 tweets. These thresholds were imposed to remove both micro-celebrities, whose profile picture might not be representative, and inactive users, who might not have a profile picture at all. For each county, up to 100 profiles were retrieved in (default) descending order of the number of followers. For each user found, we then used the Twitter REST API\footnote{\url{https://dev.twitter.com/rest/reference/get/users/lookup}} to obtain the most recent user profile information. Out of the 111 counties\footnote{Five counties with an ambiguous name were removed as explained above.}, 95 had at least one user with valid data. Details about the data size are shown in Table~\ref{tab:profiles_found}.

Note that higher obesity areas generally correspond to more rural areas with counties with smaller populations. Out of our 111 counties, the 39 low obesity counties had an average population of 185,213, whereas it was only 28,224 for the 72 high obesity counties. This means that despite the lower number of counties we actually have \emph{more} data for the low obesity counties.

\begin{figure*}[ht]
\centering
\includegraphics[width=\linewidth]{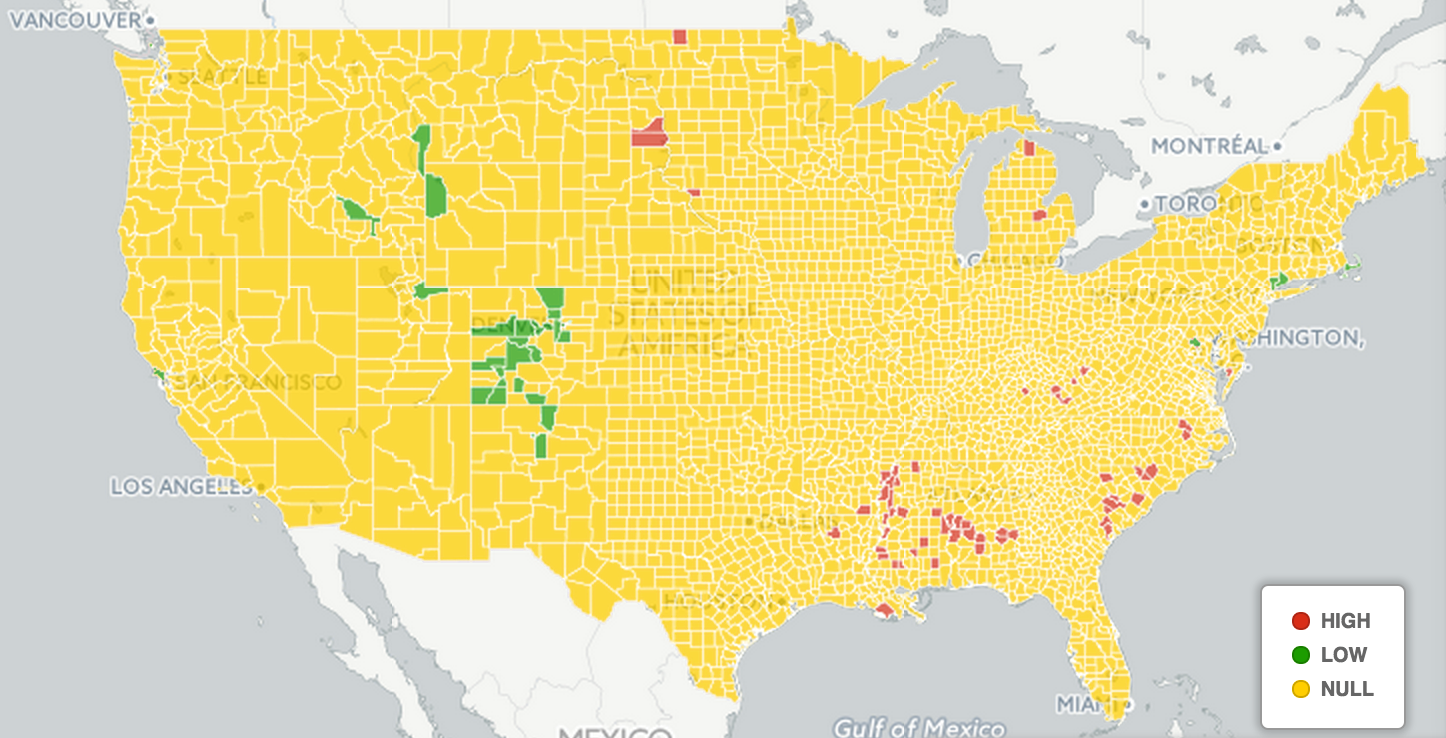}
\caption{Geographic distribution of the 1,339 users who had a valid profile picture across 83 counties. Counties in our high (low) obesity group are colored in red (green). North Slope County in Alaska in the high obesity group is not shown.}\label{fig:counties}
\end{figure*}

The most obese counties tend to be in the South whereas the least obese are between Colorado and Idaho, as well as on the East coast. Figure~\ref{fig:counties} shows the geographic distribution for the 83 counties with at least one user with a valid profile picture, as determined by the labeling task in Section~\ref{sec:is_profile}.

\section{Labeling ``Is It a Profile Picture?''}\label{sec:is_profile}

Before passing pictures to the crowd for our main ``is overweight'' labeling, we first used crowdsourcing platform CrowdFlower\footnote{\url{http://www.crowdflower.com/}} to remove all non-profile pictures. Concretely, we asked crowd workers to remove all pictures that did not satisfy all of the following conditions: (i) be an actual photo and not a drawing, (ii) show an unobstructed face, (iii) show only a single person, (iv) not show a celebrity, and (v) must not be a collage of several images. The task was a fairly easy one, with the inter-judge agreement at 94\% among the sets of three workers for each picture. 


Note that we could have potentially used face detection software instead of asking crowd workers.\footnote{See \url{http://www.faceplusplus.com/demo-detect/} for a nice online demo.} However, due to the size of our data ($<4k$ images) and its diversity in terms of lighting, race, and picture angle we preferred to have the extra accuracy that crowd-labeling provides. It also gave us the possibility to \emph{define} what we would consider acceptable or not.

\begin{table}[t]
\centering
\caption{Statistic for the Twitter profile data.\label{tab:profiles_found}}{
\begin{tabular}{ccc}
                 & High Obes. & Low Obes.  \\ \hline
\multicolumn{3}{c}{Before Screening for Pictures} \\
\# counties w.\ profiles & 60 & 35 \\
\# profiles & 1,020 & 1,766 \\\hline
\multicolumn{3}{c}{After Screening for Pictures} \\
\# counties w.\ profiles & 53  & 30 \\
\# profiles & 590 & 749 \\
\end{tabular}}
\end{table}

\section{Labeling ``Is Overweight?''}\label{sec:is_overweight}

\begin{figure}[!ht]
\centering
\includegraphics[width=0.84\columnwidth]{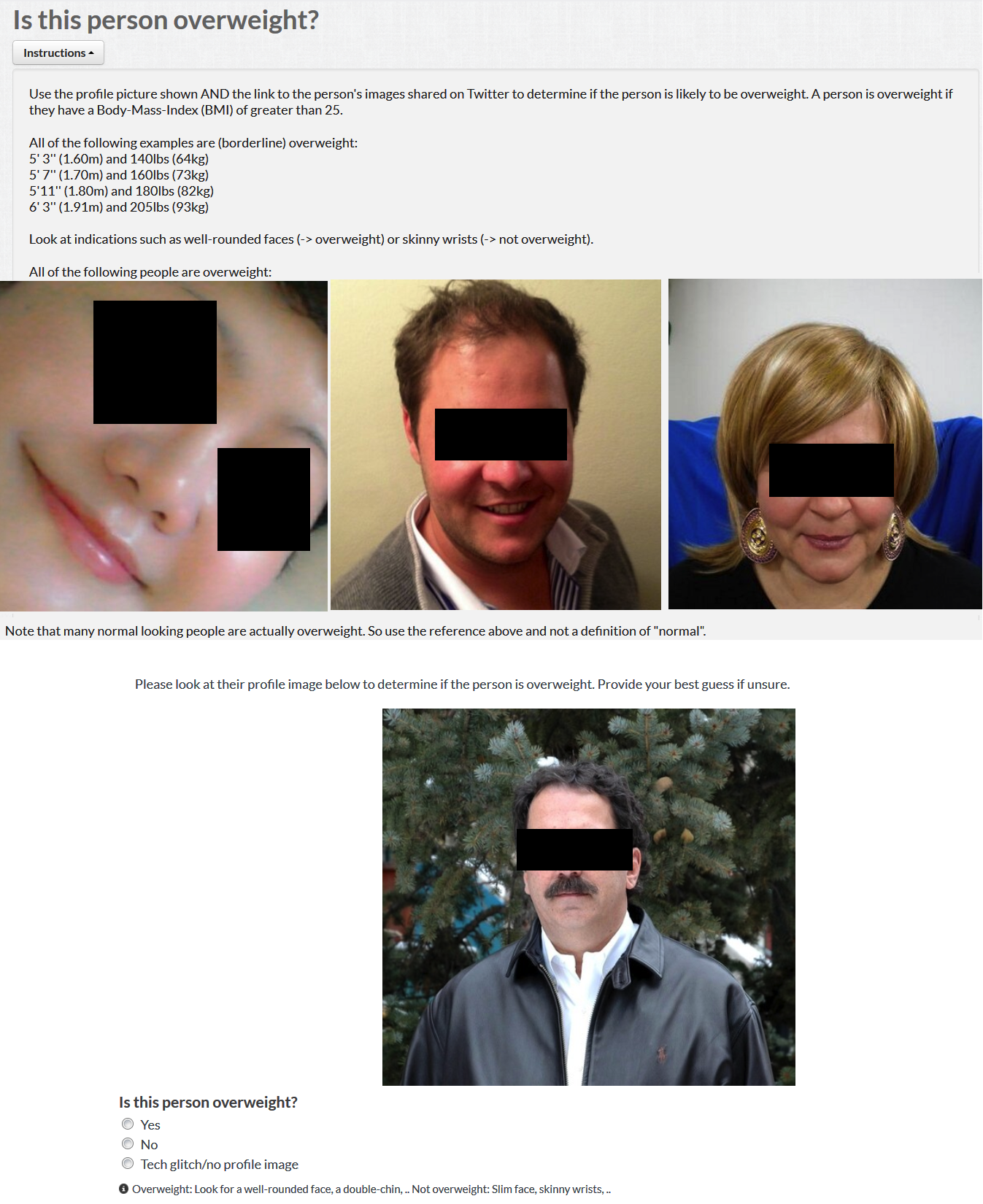}
\caption{Screenshot of the body weight labeling task done by Crowdflower workers. Example overweight pictures are shown on top as part of the instructions. 
The workers were shown the un-blackened pictures.}\label{fig:crowdflower:overweight}
\end{figure}

Judgement of whether a picture ``is overweight'' is a rather subjective task. Previous work Abbar \etal{} found that virtually no user was labeled as ``obese'' by the crowd \cite{abbaretal15chi}. We build on their experience and only ask if a user is \emph{overweight}, but not necessarily obese. Furthermore, we provide a number of example photos to illustrate that ``normal'' does not mean ``not overweight''. Figure~\ref{fig:crowdflower:overweight} shows the interface shown to the crowd workers. As we considered this task to be fairly hard and subjective, we asked for five independent judgments for each task. The inter-judge agreement was still surprisingly high at 89\% and more than 30\% of profile photos were labeled as overweight.

The CF confidence, which incorporates both the label distribution and a worker's trust score, was highest ($0.88$) for users labeled as not overweight in low obesity areas. It was lowest ($0.81$) for users labeled as overweight in low obesity areas. The difference is small though and it had no impact on our analysis whether to weight user labels by their confidence score and, for simplicity, we treated all assigned labels equally.

\section{User analysis}\label{sec:results}

\subsection{Match With Offline Data}\label{sec:offline_match}
Encouraged by the high inter-annotator agreement and the relatively high fraction of users labeled as obese, we wanted to see if the online Twitter profiles reflect the offline differences in obesity levels.

\begin{table}[ht]
\centering
\caption{Obesity labels broken down by low and high obesity counties. Eight cases of pictures not being accessible were removed for this analysis. Fisher's two-tailed exact test gave a significance of $<.0001$ in both cases, indicating a non-random relationship between the label assigned and the obesity level in the user's area.\label{tab:results_basic}}{
\begin{tabular}{ccc}
                 & High Obesity & Low Obesity  \\ \hline
\multicolumn{3}{c}{Is profile picture?} \\
yes & 590  (58\%) & 749 (43\%) \\
 no & 428 (42\%)  & 1,012 (57\%)\\\hline
\multicolumn{3}{c}{Looks overweight?} \\
yes & 251 (43\%)  & 227 (30\%) \\
 no & 339 (57\%)  & 521 (70\%) \\ 
\end{tabular}}
\end{table}

The top part of Table~\ref{tab:results_basic} shows that, surprisingly, the fraction of profiles with a valid profile picture was \emph{higher} in areas with higher obesity rates. To us this was surprising as we were expecting potential indications of ``shyness'' where people would hesitate to post profile pictures. One potential explanation could be due to the social acceptance of being overweight in high obesity counties (see \cite{kuebleretal13pone} and our Discussion later). Another reason could be a higher fraction of commercial accounts in low obesity areas, as we did not pre-filter for personal accounts. The analysis in the Obesity and Popularity Section later also indicates a higher level of Twitter usage in low obesity areas, making this indeed a viable explanation.

The bottom part of Table~\ref{tab:results_basic} shows that the fraction of profile pictures labeled as ``is overweight'' followed our expectation and was higher in areas with elevated obesity rates. This is a first indication that profile pictures contain obesity-related information that could be useful for public health analysis.

\subsection{Description of Users}\label{sec:user_description}

\begin{figure}[!th]
\centering
\vspace{-0.7cm}
\subfloat[high -- not overweight]{
  \includegraphics[width=4cm,trim={2.0cm 2.0cm 2.0cm 2.0cm},clip]{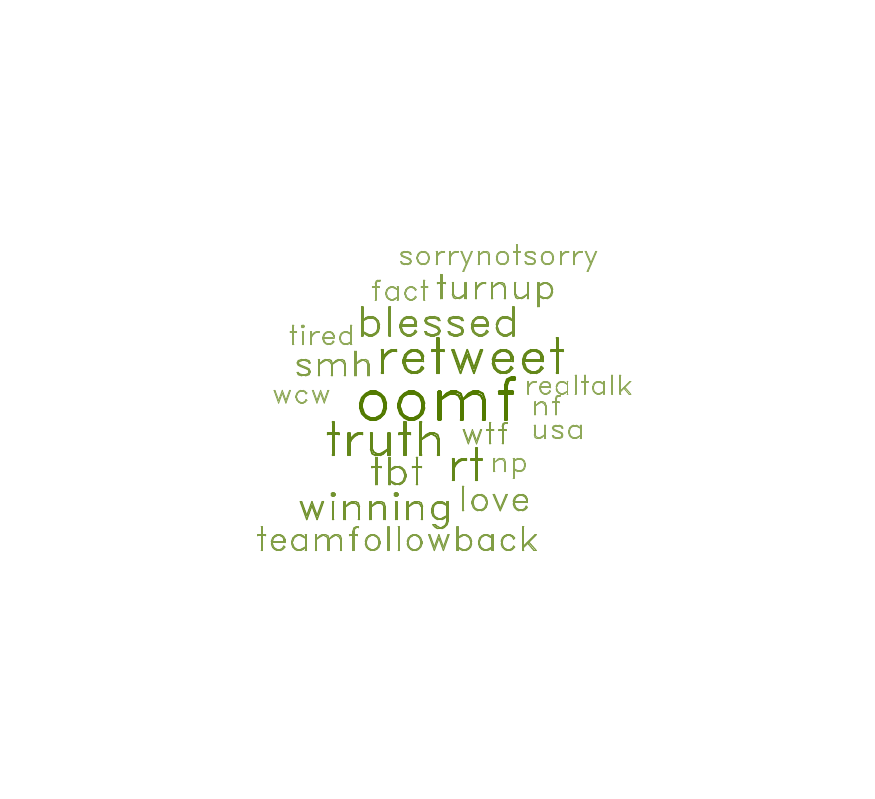}
}
\subfloat[high -- is overweight]{
  \includegraphics[width=4cm,trim={1.6cm 1.6cm 1.6cm 1.6cm},clip]{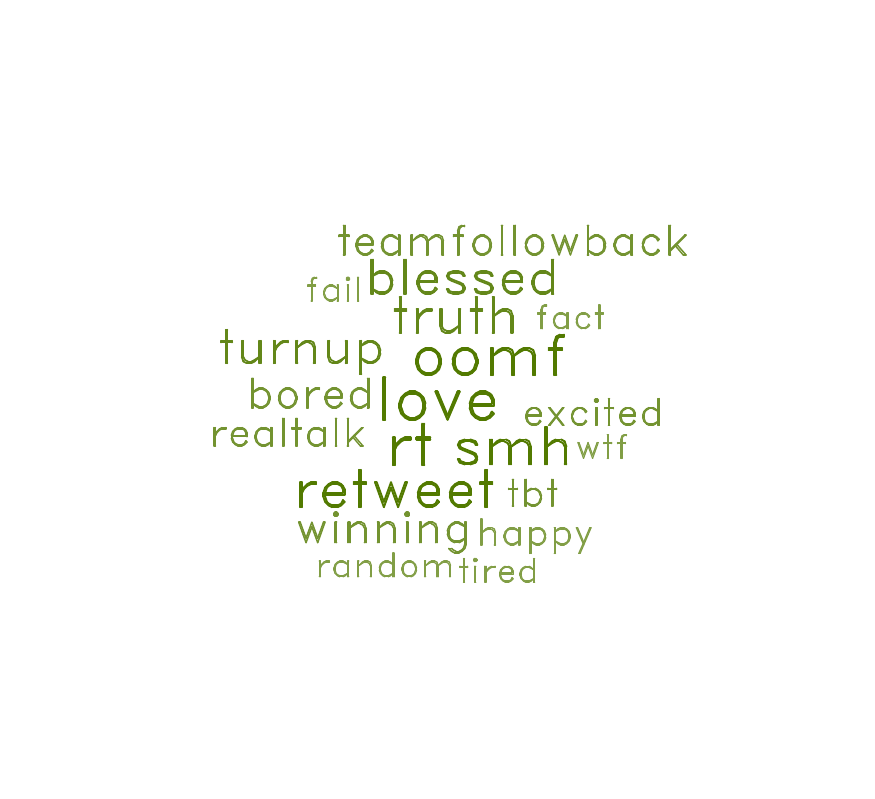}
}\\
\subfloat[low -- not overweight]{
  \includegraphics[width=4cm,trim={2.2cm 2.2cm 2.2cm 2.2cm},clip]{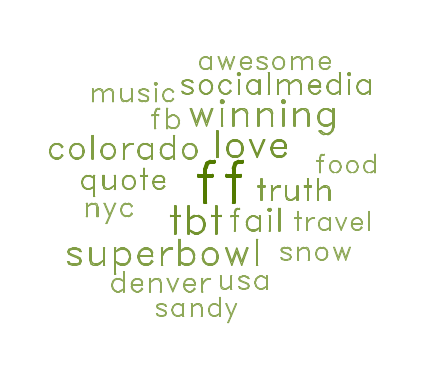}
}
\subfloat[low -- is overweight]{
  \includegraphics[width=4cm,trim={2.0cm 2.0cm 2.0cm 2.0cm},clip]{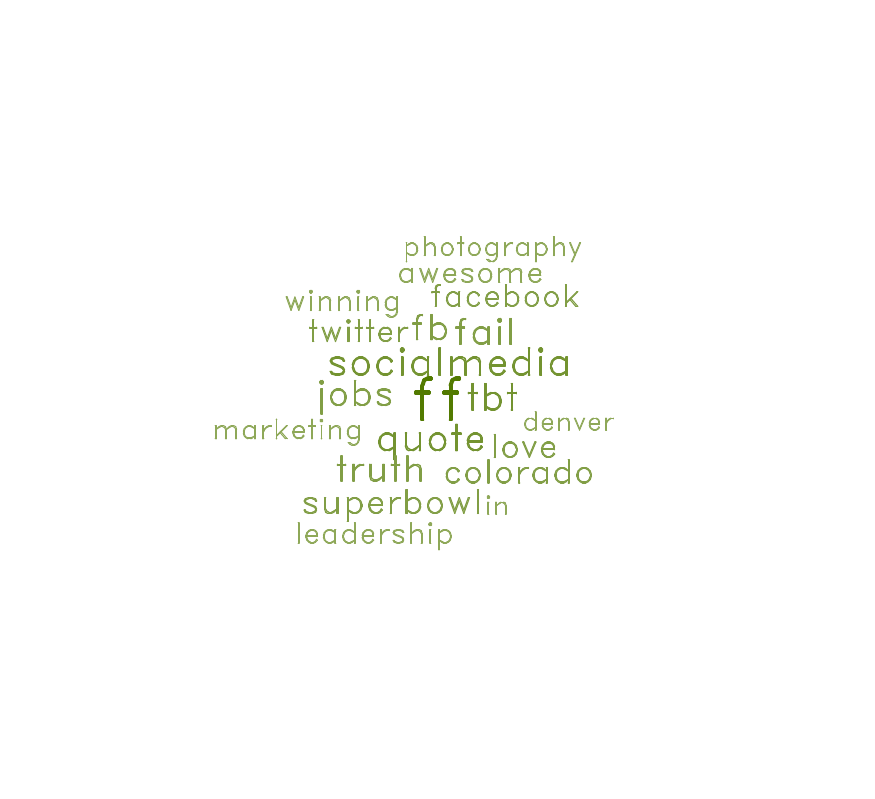}
}
\caption{\#hashtags used by different user groups. ``high''/``low'' refers to the county a user is in, ``is/not overweight'' refers to the individual's label. }\label{fig:tagclouds}
\end{figure}

As a first qualitative analysis of how overweight users differ in their behavior on Twitter from non-overweight users, we looked at the \#hashtags they used (Figure~\ref{fig:tagclouds}). 
Here, we separate low vs.\ high obesity counties and ``is overweight'' vs.\ ``not overweight'' user labels.

The differences between the areas (top vs.\ bottom set of tag clouds) are more pronounced than the differences in the individual labels (left vs.\ right set of tag clouds). For example, Twitter conventions such as \#ff for ``Follow Friday'' are more used in low obesity areas, which also have more followers and tweets (see Table~\ref{tab:followers}). 
The hashtags \#bored and \#fail occur in the top 20 of the overweight users in high obesity areas, but not in the top 20 of their non-overweight counter-parts. For the low obesity areas, the reversal in relative usage of \#winning vs.\#fail stands out, with \#fail not making the top 20 for the non-overweight users. Both sets of examples hint at a potential difference in the outlook on life which, in turn, could be linked to differences in mental health. Applying existing methods \cite{balanidechoudhury15chi} for tracking a person's mood and well-being through social media to our data set could be a promising direction for future work.


\subsection{Obesity and Popularity}\label{sec:obesity_popularity}

Next, we check for signs of weight-based popularity discrimination on social media. Table~\ref{tab:followers} shows that users labeled as ``is overweight'' have fewer followers than their ``not overweight'' counterparts, regardless of whether their location is in a high or low obesity area. At the same time they have fewer tweets, again regardless of what type of area they are in. The differences are more pronounced in high obesity areas.

\begin{table}[!ht]
\centering
\caption{Statistics for the average (median in parentheses) number of followers and tweets for high or low obesity areas, and users labeled as ``not overweight'' or ``is overweight''.\label{tab:followers}}{
\begin{tabular}{clcc}
& & Not overweight & Is overweight \\ \hline
\multirow{2}{*}{\begin{sideways}high\end{sideways}} &\# followers & 255 (199)  & 192 (113) \\ 
&\# tweets & 1,248 (684) & 965 (413) \\ \hline
\multirow{2}{*}{\begin{sideways}low\end{sideways}} &\# followers & 468 (409)  & 417 (309) \\ 
&\# tweets & 1,264 (802) & 1,117 (664) \\
\end{tabular}}
\end{table}

As the number of followers and the number of tweets is linked and users who rarely tweet typically do not attract a large followership, we built regression models both with and without the individual label included. The models tried to predict a user's number of followers based on (i) a constant intercept, (ii) their number of tweets, and, for one set of experiments, (iii) their ``is overweight?'' label. Separate models were built for the low and high obesity areas as Table~\ref{tab:followers} indicated different dynamics being at play. The results obtained for the four models are shown in Table~\ref{tab:obesity_popularity}.

\begin{table}[ht]
\centering
\caption{Linear regression predicting the number of followers. The top model uses only the number of tweets, the bottom also incorporates the ``is overweight'' label. $p<0.001 = ^{***}$, $p<0.01 = ^{**}$.\label{tab:obesity_popularity}}{
\setlength{\tabcolsep}{1pt}
\begin{tabular}{lrlrl}
   &  \multicolumn{2}{c}{Low obesity} \hspace{0.5cm}& \multicolumn{2}{c}{High obesity}\\ \hline
   Intercept & 312 & $^{***}$ & 133 & $^{***}$ \\
   \# tweets& 0.12 & $^{***}$ & 0.08 & $^{***}$\\
 $R^2$ & 0.16 &  & 0.28  & \\\hline
 Intercept &      323 & $^{***}$& 151 & $^{***}$ \\
 \# tweets &    0.11 & $^{***}$ & 0.08 & $^{***}$\\
 is overweight?\hspace{0.7cm} & -35  & & -40 & $^{**}$\\
 $R^2$ &  0.17  & & 0.29  & \\ \hline
\end{tabular}}
\end{table}

Comparing the top and bottom models of Table~\ref{tab:obesity_popularity}, one can observe a tiny increase in $R^2$. For the high obesity areas, the ``is overweight'' label is significant at $p<0.01$. This hints at the possibility that it is more difficult for overweight people to gain a large followership. Based on the model coefficients, an overweight user is likely to have $\sim$40 followers fewer in high obesity areas. Of course, the underlying causal mechanism can also be related to differences in the topics discussed (see previous section) and it is not clear if the profile picture itself plays a role. 
Note that this type of analysis would not be possible using only county-level ``ground truth'' data as, for example, done in \cite{abbaretal15chi,culotta14chi}.

\section{Discussion}\label{sec:discussion}


Note that being overweight is actually the \emph{norm} in the U.S.\ which complicates our labeling task. 69\% of U.S.\ adults are overweight\footnote{\url{http://www.cdc.gov/nchs/fastats/obesity-overweight.htm}}, significantly more than the even the 44\% found in our data for the high obesity counties. The skewed perception of ``normal = not overweight'' could be one reason for the comparable low percentage of users labeled as overweight. In future work we plan to include characteristics of the crowd workers into the analysis. For example, do crowd workers from areas with higher obesity rates and, hence, potentially a different perception of ``normality'' have a higher tendency to label users as non-overweight?

A human bias towards falsely equating overweight with abnormal could potentially be overcome by using machine-learned computer vision techniques such as those explored in \cite{wenguo13ivc}. It is, however, not clear if such methods will work for the noise and diversity that is inherent to social media images. We plan to experiment with training and testing such methods on profile pictures in future work.

Apart from depending on images, another option to obtain ground truth on a user's weight could come from ``quantified self'' users \cite{wangetal16dh}.
Companies such as Withings produce ``smart scales'' that can automatically tweet a user's weight.\footnote{Product description: \url{http://www.withings.com/eu/smart-body-analyzer.html}} Note that these tweets\footnote{Example tweets: \url{https://twitter.com/search?q=withings}} lack a user's height, though this additional information, which is required to reason about obesity, could potentially be present in a user's profile page on sites used by quantified self users.

Text-based approaches could offer another method to infer a user's weight status. For example, we discovered through manual inspection that certain Twitter hashtag such as \#FatGuyProblems or \#FatPride are generally indicative of the user being overweight. However, the level of noise is higher than in our approach and the number of user providing such explicit signals is also far lower than the number of users with a valid profile photo.

Another generic approach to infer labels for a user in a social network is to incorporate knowledge and labels from their social connections. For virtually all variables of interest, there are strong clustering effects where ``birds of a feather flock together'' \cite{alzamaletal12icwsm}. For the type of analysis we would like to do, such an approach would be counter-productive though, as it could create systematic biases, basically pre-imposing effects that one might want to measure, such as the social influence in obesity propagation \cite{christakisfowler07nejm}.

We found it to be of utmost importance to make explicit to crowd workers that overweight is not ``abnormal''. The perception of what is normal also has implications for mental health. Using data from Yahoo Answers, Kuebler \etal{}\ \cite{kuebleretal13pone} found evidence that ``obese people residing in counties with higher levels of BMI may have better physical and mental health than obese people living in counties with lower levels of BMI''. This also relates to the ``contact hypothesis'' \cite{allport54perseus} stating that discrimination decreases as the discriminators are more exposed to direct contact with the discriminated group. Having individual-level labels could shed light on such questions.


When using social media for studies in health or on other topics it is important to realize that the collected data is \emph{not} representative in the sense of being a uniform sample of the population.  
This, however, does not preempt the usage of this data for population-level studies if proper weighting schemes are used \cite{zagheniweber15ijm}.



\section{Conclusion}

Our paper shows the feasibility to obtain a new type of health-related user labels, laying the ground-work for further public health studies using social media data. More work is however needed to reduce a bias of human annotators to falsely equate ``overweight'' with ``abnormal''. Machine-learned approaches could help to address this issue.


\bibliographystyle{SIGCHI-Reference-Format}
\bibliography{image_weight}

\end{document}